\documentclass[9pt,twocolumn,twoside]{pnas-new}

\templatetype{pnasresearcharticle} 

\usepackage{algcompatible}
\usepackage{multirow}

\algblock{If}{EndIf}
\algcblock[If]{If}{ElsIf}{EndIf}
\algcblock{If}{Else}{EndIf}
\algcblockdefx[Strange]{}{Eeee}{Oooo}
[1]{\textbf{Input} #1}
[1]{\textbf{Output} #1}

\begin{document}

\title{Multidisciplinary learning through collective performance favors decentralization}

\author[a,1]{John Meluso}
\author[a,b]{Laurent H\'{e}bert-Dufresne}

\affil[a]{Vermont Complex Systems Center, University of Vermont, Burlington, VT 05405}
\affil[b]{Department of Computer Science, University of Vermont, Burlington, VT 05405}

\leadauthor{Meluso \& H\'{e}bert-Dufresne}

\significancestatement{
Like chefs at a fast-moving restaurant or engineers in a multidisciplinary project, team members often complete separate, interrelated subsets of larger tasks with limited insight into the work of others.
These contexts make it difficult for individuals to assess the value of their own contribution to the collective work.
Our work shows that despite this obstacle, individuals can still learn from their neighbors when neighbors' actions influence collective outcomes.
Though the effects are modest, we found that teams with more interactions between members perform better when refining their work while teams with fewer interactions perform better when innovating.
We also found that across 34 tasks with diverse qualities, teams that decentralize coordination responsibilities outperform those that do not.
}

\authorcontributions{JM \& LHD contributed conceptualization, methodology, project administration, validation, and writing (review \& editing). JM contributed data curation, formal analysis, investigation, software, visualization, and writing (original draft). LHD contributed funding acquisition, resources, and supervision.}
\authordeclaration{JM \& LHD are supported by Google Open Source under the Open Source Complex Ecosystems And Networks (OCEAN) project, and by the Sloan Foundation through the Vermont Research Open Source Program Office (VERSO). LHD is also supported by the National Science Foundation award EPS-2019470.}
\correspondingauthor{\textsuperscript{1}To whom correspondence should be addressed. E-mail: john.meluso@uvm.edu}


\begin{abstract}
Many models of learning in teams assume that team members can share solutions or learn concurrently.
However, these assumptions break down in multidisciplinary teams where team members often complete distinct, interrelated pieces of larger tasks.
Such contexts make it difficult for individuals to separate the performance effects of their own actions from the actions of interacting neighbors.
In this work, we show that individuals can overcome this challenge by learning from network neighbors through mediating artifacts (like collective performance assessments).
When neighbors' actions influence collective outcomes, teams with different networks perform relatively similarly to one another.
However, varying a team's network can affect performance on tasks that weight individuals' contributions by network properties.
Consequently, when individuals innovate (through ``exploring'' searches), dense networks hurt performance slightly by increasing uncertainty.
In contrast, dense networks moderately help performance when individuals refine their work (through ``exploiting'' searches) by efficiently finding local optima.
We also find that decentralization improves team performance across a battery of 34 tasks.
Our results offer design principles for multidisciplinary teams within which other forms of learning prove more difficult.
\end{abstract}

\dates{This manuscript was compiled on \today}
\doi{\url{www.pnas.org/cgi/doi/10.1073/pnas.2303568120}}

\maketitle
\thispagestyle{firststyle}
\ifthenelse{\boolean{shortarticle}}{\ifthenelse{\boolean{singlecolumn}}{\abscontentformatted}{\abscontent}}{}

\firstpage[11]{4}

\dropcap{M}ultidisciplinary teams play crucial roles in academia, industry, and government \cite{NAP2014Convergence}. Scientists from different fields collaborate to solve complex problems \cite{NAP2014Convergence}. Chefs train as sauciers or pastry chefs, then combine their products into the same meals \cite{Rambourg2013, Gregoire2016}. Engineering teams draw representatives from various backgrounds, each working on parts that integrate with the others \cite{Wertz2011, Meluso2022}. But these engines of innovation face an obstacle familiar to many: disciplinary boundaries make collaborating across fields challenging \cite{Levina2005, Cronin2007, Barley2015, Kotlarsky2015, Edmondson2018, Mattarelli2021}. How, then, do members of multidisciplinary teams learn from and with one another as they advance toward collective goals?

\begin{figure}
    \centering
    \includegraphics[width=7cm]{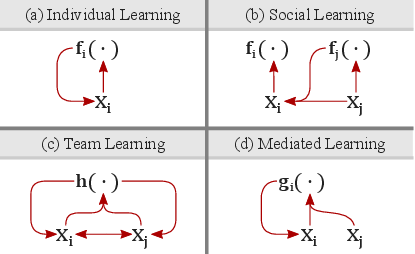}
    \caption{\textbf{Types of learning in teams.} In each panel, $i$ is some member of a team, and $j$ can be any other member of that team; $x_i$ and $x_j$ are their respective decisions. Then, $f_i$ is a task that $i$ can work on alone; $h$ is a task that an entire team works on; and $g_i$ is a task that team member $i$ works on but that is also influenced by the actions of their network neighbors $j$. (a) Individuals can learn from their own actions without outside influences. (b) Individuals can also learn by copying solutions from others they interact with. (c) Teams can learn concurrently through their actions, interactions, and feedback. (d) In multidisciplinary teams, individuals can learn through mediating artifacts, like performance assessments, that provide feedback on tasks that multiple individuals contribute to. This does not necessarily require social or team learning because knowledge boundaries between disciplines and inseparable contributions to artifacts tend to impede these forms of learning between disciplines.}
    \label{fig:learning_types}
\end{figure}

Prior work on learning in teams would suggest two possibilities. The first treats individuals' actions as ``copyable'' \cite[e.g.][]{Lazer2007, Lorenz2011, Hong2012, Mason2012, Barkoczi2016, Bernstein2018, Toyokawa2019, Brackbill2020, Miu2020, Yang2021}, like learning sales strategies from a fellow salesperson or learning the steps to diagnose an illness from a senior physician. Here, people learn through feedback from their own actions (\textit{individual learning}, Fig. \ref{fig:learning_types}a) and by copying techniques from those around them (\textit{vicarious} or \textit{social learning}, Fig. \ref{fig:learning_types}b) \cite{Bandura1989, Grusec1992, Bandura2002, Kendal2018, Guillo2020}. Tasks \cite{Rivkin2007}, networks \cite{Mason2012}, and learning processes \cite{Barkoczi2016} all affect these types of learning.

In the second perspective, team members work together concurrently or ``transactively'' \cite[e.g.][]{Lewis2005, Woolley2010, Lewis2011, Engel2014, Barlow2016, McHugh2016, Amelkin2018, Almaatouq2021, Riedl2021, Ostrowski2022, Woolley2022}, like software engineers developing a new app together \cite{Faraj2000} or a medical team treating patients in a busy emergency room \cite{Valentine2014}. These teams learn from performance feedback while they dynamically encode, store, retrieve, and integrate different kinds of knowledge (\textit{team learning}, Fig. \ref{fig:learning_types}c) \cite{Lewis2005, Lewis2011}. Such teams are often described in terms of their collective intelligence, a validated quality describing a group's ability to perform a wide variety of tasks \cite{Woolley2010, Riedl2021} which likewise depends on tasks \cite{Graf-Drasch2021}, interaction patterns \cite{Woolley2010, Bernstein2018, Woolley2022} and skills \cite{Riedl2021}.

However, these perspectives prove limited within multidisciplinary teams. Disciplinary boundaries often impede individuals from copying each other's actions or working closely together \cite{Levina2005, Cronin2007, Barley2015, Kotlarsky2015, Edmondson2018, Mattarelli2021}. Instead, they divide tasks into interdependent subtasks and work in parallel \cite{Kiggundu1983, Saavedra1993, Wageman1995}. Of course, members of these teams still interact with one another, affect each other's outcomes, and the team's collective outcomes. This gives rise to the ``performance non-separability problem'' \cite{Alchian1972, Uribe2022}: a team's collective performance obfuscates how individuals' actions affect shared outcomes, making it hard to figure out the effect of each change. Learning becomes even more challenging when many changes happen at the same time (they may not work together) or individuals have limited visibility into each other’s actions \cite{Postrel2002, Stasser2003}. So while extant forms of learning remain possible, multidisciplinary settings do not lend themselves to individual learning, social learning within the team, or team learning.

\begin{figure*}
    \centering
    \includegraphics[width=15.5cm]{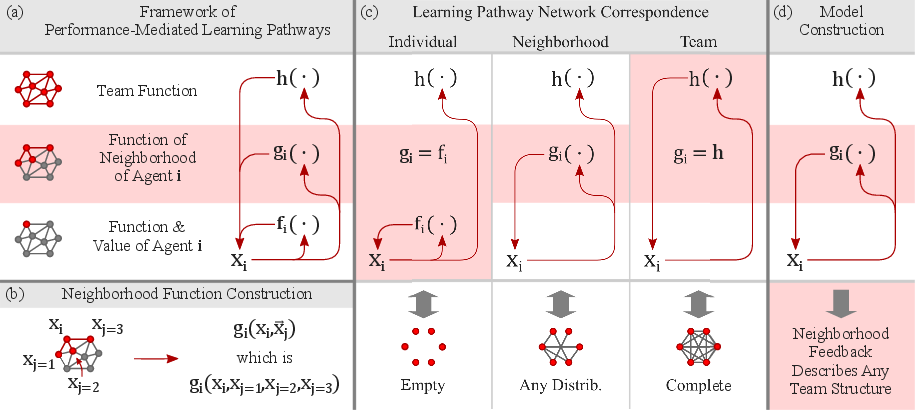}
    \caption{\textbf{Framework of performance-mediated learning pathways describes different networks, and comprehensive model construction.} (a) In the framework, individuals $i$ can learn from team performance $h$ or their own performance $f_i$, but in multidisciplinary teams they often have to learn from a combined ``neighborhood'' performance $g_i$. (b) In this context, the individual's actions $x_i$ and their neighbors' actions $\Vec{x}_j$ affect the agent's performance on a subtask $g_i$ of the team's overall task $h$. This makes it difficult for individuals to tell ``how good'' their actions are. (c) These neighborhoods of dependence can describe any performance-mediated learning pathway, from individual learning (an empty graph) to the performance feedback paths of team learning (a complete graph) and any distribution in between. (d) Our model enables us to explore different networks by giving each individual $i$ feedback from their neighborhood $g_i$. Separately, our model assesses the team's overall performance $h$ as a function of all agents actions $\Vec{x}$ .}
    \label{fig:function_relationships}
\end{figure*}

Fortunately, another learning pathway exists: \textit{mediated learning}. Here, individuals learn from themselves and others through artifacts \cite[Fig. \ref{fig:learning_types}d, ][]{Kozulin2003, Guillo2020}. For example, members of multidisciplinary teams can learn about their performance and their teammates’ performance through changes in the quality of the collective output of the team. Consider how ants learn where food is by following chemical trails left by other ants \cite{Jackson2006}. People learn from one another through symbolic objects they create, like books, websites, and performance assessments \cite{Kozulin2003, Guillo2020, Uribe2022}. Team members learn from one another through boundary objects, like sketches and databases \cite{Carlile2002, Majchrzak2011, Nicolini2011}. Such mediated interactions can significantly affect the outcomes of microbes \cite{Xiao2022}, ecological webs \cite{Ohgushi2005}, social behavior spread \cite{Xie2022}, education \cite{Henrie2015}, and innovation \cite{Ahuja2000}, even producing larger effects than unmediated interactions \cite{Xie2022}. Mediated learning is not mutually exclusive of the other learning types as individuals and teams can learn through artifacts as well.

Mediated learning has been observed in many settings, but to our knowledge, the performance effects of this subtle mechanism have yet to be examined systematically in multidisciplinary teams. In particular, network properties produce varied performance effects depending on the type of learning \cite{Barkoczi2016, Xie2022} which makes systematic analysis of network properties crucial to understanding how mediated learning affects team outcomes. To that end, this work addresses the question: How do network properties affect multidisciplinary team performance through mediated learning?

We begin by introducing neighborhood task performance, which formalizes the networked nature of mediated learning pathways and distinguishes them from other learning pathways. Through this concept, we investigate how network density and other properties affect team performance, showing three results. First, mediated learning can either help or hurt performance depending on the task and team network structure. Then, we show that densely-connected teams tend to perform better when ``exploiting'' (or refining) solutions, while sparsely-connected teams perform better when ``exploring'' (or innovating) new solutions---a finding that echoes other types of learning \cite{Lazer2007, Shore2015, Derex2016}. But contrary to other types of learning \cite{Mell2014, Kocak2022}, we show that \textit{decentralized} teams see greater performance benefits than other structures across a diverse set of tasks. This benefit likely arises because decentralization balances the benefits of dense and sparse teams on tasks with different qualities. We close with a discussion of implications for designing teams and limitations.

\section*{Model description}

We constructed a model that represents mediated learning in multidisciplinary teams. To achieve this, we simulated performance-mediated feedback networks, team members as computational agents, and representative tasks while minimizing other forms of learning and variables with established effects. In brief, each team member independently selects actions. Individuals' actions affect performance on their own subtask, but also affect their neighbors' subtask performances and the team's overall task performance. This means individuals learn from a ``neighborhood'' performance shaped by their own actions and the actions of their network neighbors.

\subsection*{Performance-mediated feedback framework}
\label{subsec:model-performance-feedback-framework}

The performance non-separability problem exemplifies mediated learning in multidisciplinary teams. Here, team members perform actions that affect each other's work. In other words, individuals' performances depend on the actions of other individuals. This yields ties in a network consisting of performance dependencies between individuals. Then, when individuals receive feedback from those confounded performance assessments (as in Fig. \ref{fig:learning_types}d), these action-performance-feedback sequences create a network of pathways for learning. Rather than being individual learning (learning from feedback on one's own actions in isolation) or social learning \cite[learning by observing and copying others actions, as in][]{Lazer2007}, this is mediated learning because individuals learn through mutually-constructed artifacts---collective performance assessments that result from multiple individuals' actions. Thus, this scenario describes performance-mediated learning in multidisciplinary teams.

In our model, then, computational agents work on a shared team-level fitness landscape representing a task (called a ``task'' hereafter) over time. Each team has $n$ agents, and each agent $i\in\{1,\ldots,n\}$ has a value $x_i\in[0,1]$ that represents its decisions on how to contribute to the task. Like others \cite{Hong2001, Lazer2007, Rivkin2007, Mason2012}, we assume that a group's performance on a task can be represented as a function $h(\Vec{x})$ of the agents' decisions. The context we study in this work means we cannot assume that individuals only receive individual feedback via some $f_i(x_i)$, as with individual learning, or team feedback via $h(\Vec{x})$, as with team learning (see Fig. \ref{fig:function_relationships}a). Instead, agent $i$ owns and works on a subtask $g_i$ of the team's overall task $h$. The subtask is affected by the agent's own actions $x_i$, but not \textit{only} those actions. As in the performance non-separability problem, agent $i$ does not have complete control over its outcome because the actions $\Vec{x}_j$ of some set of other agents $j\in J_i=\{1,\ldots,k_i\}$ also affect agent $i$'s performance on $g_i$.

This yields what we call a neighborhood function $g_i(x_i, \Vec{x}_j)$ for each agent $i$. Neighborhood functions describe an agent's subtask performance as a function of the agent's own actions and the actions of other agents that affect it (Fig. \ref{fig:function_relationships}b). This makes neighborhood functions $k_i+1$ dimensional, with one dimension for $i$ and one for each $j\in J_i$. Neighborhoods are not sub\textit{teams}, though, as there is no coordination between agents and agents' actions can affect multiple other agents. So each neighborhood function $g_i(x_i, \Vec{x}_j)$ describes agent $i$'s subtask contribution to the team's performance $h$ while acknowledging agent $i$'s limited control over its outcome.

Collectively, a team's neighborhood functions correspond to the team's mediated learning network with nodes $i$ and network neighbors $j\in J_i$.\footnote{More completely, $g_i(x_i, \Vec{x}_j)$ describes the directed edges of a network of mediated interactions, where $x_i$ indicates the existence of a self-edge and each $x_j$ indicates the existence of a directed edge from $j$ to $i$. For simplicity, we assume that all edges in this network are reciprocated, making a network of undirected edges sufficient for this work.} This multi-level framework can describe any performance-mediated learning pathway (Fig. \ref{fig:function_relationships}c). At one extreme, agents only affect themselves (representing individual learning, or an empty graph). As we add interactions, we can construct any undirected network distribution by adding interactions to neighborhood functions $\Vec{g}$ up to and including team learning (a complete graph) in which every agent affects every other agent. In this work, we vary the neighborhood functions to study the effects of mediated learning networks on team performance (Fig. \ref{fig:function_relationships}d).

Individuals often have different goals from the team \cite{Scott2007}. We could choose any task $g_i$ for each agent to work on, regardless of the task $h$ on which the team is evaluated. In this work, we assume agents work toward the same goal as the team. We do this by using tasks that are commutative, normalized, and scalable to any number of dimensions. This lets us use the same task type for both neighborhood subtasks $g_i$ and the team task $h$ because the operand corresponding to each agent $i$ retains the same form and function in every subtask in which it exists and in the team task (see SI Appendix S1 for details).

Put another way, $g_i$ captures ``how well'' agent $i$ thinks their action contributes to the team's overall task $h$. In many cases, agent $i$ cannot tell how their own decision $x_i$ led to their own outcome $g_i$ (let alone the team's outcome $h$) because the $\Vec{x}_j$ decisions of the other agents $j\in J_i$ obscure how agent $i$'s actions affected collective outcomes. For example, while one agent could make a decision that would perform well in isolation, adjacent agents may be negatively affected by the decision. This can occur because agents work on different subtasks from one another (again, neighborhoods are not subteams) and a decision that improves one agent's subtask performance might degrade another agent's subtask performance. On the other hand, the decision could also improve the performance of adjacent agents $j$ either by increasing $g_j$ or by opening new possibilities.

\begin{figure*}
    \centering
    \includegraphics[width=\textwidth]{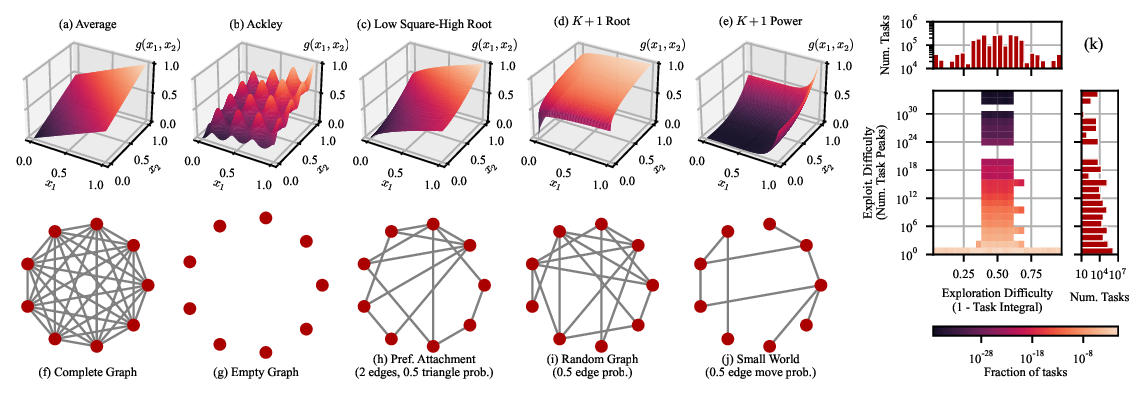}
    \caption{\textbf{Example tasks, networks, and their difficulties.} We put teams through different fitness landscapes representing 34 tasks while in each of 18 network configurations, including (a-e) the example tasks and (f-j)  networks shown here. (k) Together, the task functions and networks varied task difficulty across four measures, including for exploration and exploitation as shown here.}
    \label{fig:example_tasks_networks}
\end{figure*}

\subsection*{Agents description}
\label{subsec:model-agents}

Given this framework, we give agents four qualities. First, agents act independently and cannot coordinate future actions by sharing their intended states. We chose this because multidisciplinary teams may not be able to discern how each other's actions will affect their own outcomes. Also, sharing information with one another is part of both social and team learning which we sought to minimize. So, agents' only interactions are mediated through how their choices affect each other's outcomes as described through their neighborhood performance functions $g_i(x_i,\Vec{x}_j)$.

Second, agents are naive. Individual skill significantly affects team and task performance \cite{Amelkin2018, Riedl2021}. To minimize the effects of individual skills on team performance, agents randomly search for better values of $x_i$ once per time step. Agents do this by drawing a single random value, evaluating that value against their prior best value, and keeping the better value of the two. This random individual experimentation cannot remove \textit{all} skill as it grants agents the ability to predict how one potential value change might affect their outcome at each time step. However, this naive approach minimizes skill for our examination of mediated learning because alternative search approaches either increase individual skill (i.e. employing search strategies, multiple sampling, error checking) or necessitate adding some unmediated form of learning (i.e. coordinating team experiments) which prove challenging in multidisciplinary settings.

Third, agents are search-limited. In the tradition of March \cite{March1991} and others \cite{Gupta2006, Lazer2007, Fang2010, Mason2012}, this represents the extent to which individuals try to innovate by exploring new solutions and refine solutions by exploiting the current solution (whether by choice or constraints). When agents sample new values $x'_i$, they explore within a specified search radius $r$---so they look for a new value $x'_i \in [x_i-r,x_i+r]$, but lower-bounded by 0 if $x_i-r<0$ and upper-bounded by 1 if $x_i+r>1$. We can then examine how one's choice of search conditions affects performance, varying from local refining searches to global exploring searches.

Finally, agents are also greedy. This means that agents seek to maximize performance by accepting any new value that they expect to improve their neighborhood's performance. We chose this decision-making process for two reasons. For one, it is consistent with the incentive structures of prior lab studies and the explicit rules in simulation studies of learning in teams \cite{Lazer2007, Mason2012, Shore2015, Barkoczi2016}. Second, it is a uniform performance-maximizing rule for assessing how individuals learn through mediated artifacts. Other rules are certainly possible \cite[e.g. accepting every third value, or only improving values until they are ``good enough'',][]{Simon1991}, but such rules add a variable better explored after establishing a baseline.

\subsection*{Simulation description}
\label{subsec:model-simulation}

To summarize, every turn of the model, each agent performs a naive search for a better value of $x_i$ by selecting a random value $x'_i$ within a given search radius $\pm r$. If the new value $x'_i$ produces a better outcome given what they know from the team's past choices---if $g_i(x'_i,\Vec{x}_j)>g_i(x_i,\Vec{x}_j)$---they accept the new value by setting $x_i=x'_i$.  They do not confirm that their decision produces the intended effect (a higher-skill verification technique) but simply continue searching. Otherwise, they keep their old $x_i$ and continue searching (see Materials \& Methods for pseudocode).

Altogether, this construction allows us to systematically explore network properties while controlling for variables of known significance, including team size \cite{BouchardJr1970, Gallupe1992, Lamberson2011, DeVincenzo2017}, tasks \cite{Amelkin2018, Graf-Drasch2021}, and the balance between innovating (exploring) and refining (exploiting) their solutions \cite{March1991, Lazer2007, Fang2010, Mason2012}. We constructed teams of different sizes---4, 9, 16, and 25 agents---that span those represented in other studies \cite[e.g.][]{Lazer2007, Woolley2010, Mason2012}.

We controlled for four realistic qualities of tasks as popularized by McGrath \cite{McGrath1984} and recommended for studies on collective intelligence \cite{Woolley2010, Riedl2021} and transactive memory systems \cite{Lewis2011, Bachrach2019}. These task qualities---generate, choose, negotiate, and execute---map to four quantities that we can measure---respectively: exploration difficulty, exploitation difficulty, neighborhood alignment, and neighborhood interdependence (see SI Appendix S2 for details). We designed a battery of 34 tasks that vary in difficulty for each of these four measures to understand team performance across a diverse set of tasks, much like how studies of collective intelligence select tasks from each group \cite[e.g.][]{Woolley2010, Amelkin2018} (see Fig. \ref{fig:example_tasks_networks}a-e for sample tasks; see SI Appendix S1 for all task specifications).

We also controlled for the search radius of agents from much smaller than the smallest feature size of the tasks ($r=0.001$), through local ``exploitation'' searches just smaller than the width of the smallest valley in the task landscape ($r$=0.01), to intermediate searches that mix exploration and exploitation ($r=0.1$), and global ``exploration'' searches in which agents can always explore the entire domain ($r=1$).

For our systematic analysis of network structures, we selected 12 measures that describe different structural properties of networks related to individual connectedness, neighbors' connectedness, network efficiency, and shared connections \cite[cf.][see SI Appendix S3 for definitions]{Newman2018}. Then, we selected 18 team network configurations that vary these qualities to understand how each quality affects team performance (see Fig. \ref{fig:example_tasks_networks}f-j for sample networks; see SI Appendix S4 for network specifications). Some of the networks are common empirically like the small world \cite{Watts1998} and preferential attachment \cite{Barabasi1999}; others are random graphs to explore the effects of network density; and several are idealized or idiosyncratic graphs because they allow us to explore particular qualities. Combined, the task and network types produced varied difficulties for the teams (see Fig \ref{fig:example_tasks_networks}k for two of the four task measures).

Finally, we ran each combination of these choices 250 times to examine statistical differences between population means. All agents began from a random starting point during each run of the model and conducted a search at each of 25 time steps. The following section presents our results.

\section*{Results}
\label{sec:results}

\begin{figure*}
    \centering
    \includegraphics[width=\textwidth]{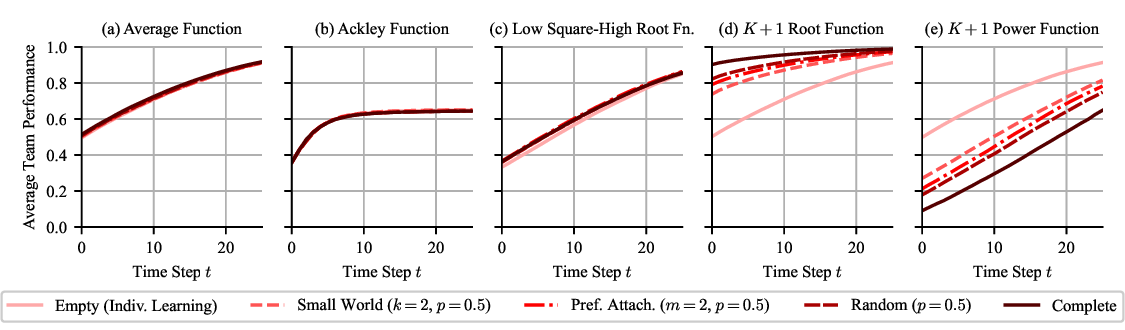}
    \caption{\textbf{Examples of team performances averaged over 250 runs for search radius {\boldmath$r=0.1$}.} (a-b) Network structure does not affect team performance on some tasks (10 of 34). However, network structure does affect performance on others (24 of 34 tasks), typically when the tasks are a function of network properties. (c) For example, on some tasks, teams with only individual learning (an empty graph) perform worse than teams with any interactions (any other graph). (d) On others, increasing network density corresponds to increased performance, (e) while others still exhibit decreased performance with greater density.}
    \label{fig:example_outcomes}
\end{figure*}

\subsection*{Mediated learning can help or hurt performance}
\label{subsec:results-tasks-favor-networks}

To answer our research question, we begin by considering how well teams perform on different tasks when we vary their interaction network's density. Intuitively, one might expect agents to affect the quality of solutions that neighboring agents find. For example, one agent might randomly start with a low-scoring value, making it likely that most values that their neighbors find improve their neighborhood performances and team performance. This implies that greater density (more interactions) might give agents the flexibility to find better solutions to their portions of the problem---even when they did poorly themselves---or conversely to constrain them when their neighbors do poorly. We examined this intuition by calculating the average team performance for each combination of team size, task, search radius, network, and point in time over the 250 runs.

The results show that density affects task performance, but only in specific cases. Fig. \ref{fig:example_outcomes} demonstrates how density can have varied effects on team performance. For some tasks, teams perform the same as one another regardless of their density (e.g. in Figs. \ref{fig:example_outcomes}a-b). Consider an intermediate search radius of $r=0.1$, for example. On 10 of the 34 tasks (29\%), the average performance of teams with any interactions (all 17 non-empty graphs) is the same as for teams with no interaction (an empty graph, only individual learning) with a 95\% confidence interval.

However, this is not always the case. Figure \ref{fig:example_outcomes}c shows a task in which teams with \textit{any} interactions between agents outperform teams with no interaction. This effect is exaggerated further in Fig. \ref{fig:example_outcomes}d, depicting a task where teams with more interaction between agents (and higher network density) find better solutions faster than those with less interaction (and lower density). Teams with interactions outperform teams with no interactions on 8 of 34 tasks (24\%), again true for all 17 non-empty graphs (95\% c.i.). In contrast, Fig. \ref{fig:example_outcomes}e shows one of the 8 tasks (24\%) in which having more interactions decreases the likelihood of finding better solutions (all 17 graphs, 95\% c.i.). Of the remaining tasks, 3 yield an advantage and 5 yield a disadvantage for varied combinations of interacting teams (see SI Appendix S5 for average performances for all tasks, search radii, and networks for teams of $n=9$).

The formulas of the tasks (see SI Appendix S1) shed light on why teams with certain networks perform differently from one another. Of the 16 tasks on which interacting teams have a distinct advantage or disadvantage, 15 of them multiply and/or exponentiate each agent's decision variable $x_i$ by the number of other agents $k_i$ that the agent interacts with. The remaining task, the minimum function, strongly depends on the number of values in its operating set which is also a function of the number of interacting neighbors. Put another way, teams with different networks tended to perform differently from one another when tasks weighted agents' contributions by a network property.

In hindsight, it may seem obvious that a team's network will affect performance when tasks depend on network properties. Nevertheless, this is realistic. For example, in a group brainstorming task (a generate task), individuals have different information depending on whether they generate ideas alone (an empty graph), with every member of a group (a complete graph), or with some subset (a distribution), with well-established repercussions for the team's performance \cite{Taylor1958, BouchardJr1970, Mullen1991, Kerr2004, Amelkin2018}. The same is true on many choose tasks \cite{Kittur2009, Shore2015, Barkoczi2016, Almaatouq2021}. Next, we explore how each network performs across our full set of tasks.

\subsection*{Networks can help or hurt performance across diverse tasks}
\label{subsec:results-networks-help-hurt}

Recent work on collective intelligence has raised the importance of understanding how well teams perform, not just on single tasks, but across sets of diverse tasks \cite[like in][]{Woolley2010, Amelkin2018, Riedl2021}. In light of this and our research question, our next analysis explores how team performance varies with network density when averaged across our 34 tasks. We calculated the average team performance for the 18 network types, four search radii, and four team sizes across all 34 tasks. As a point of comparision, we then calculated the percent difference of these values relative to a set of baseline values---the average performance of teams that only learned individually (the empty graph).

\begin{figure*}
    \centering
    \includegraphics[width=\textwidth]{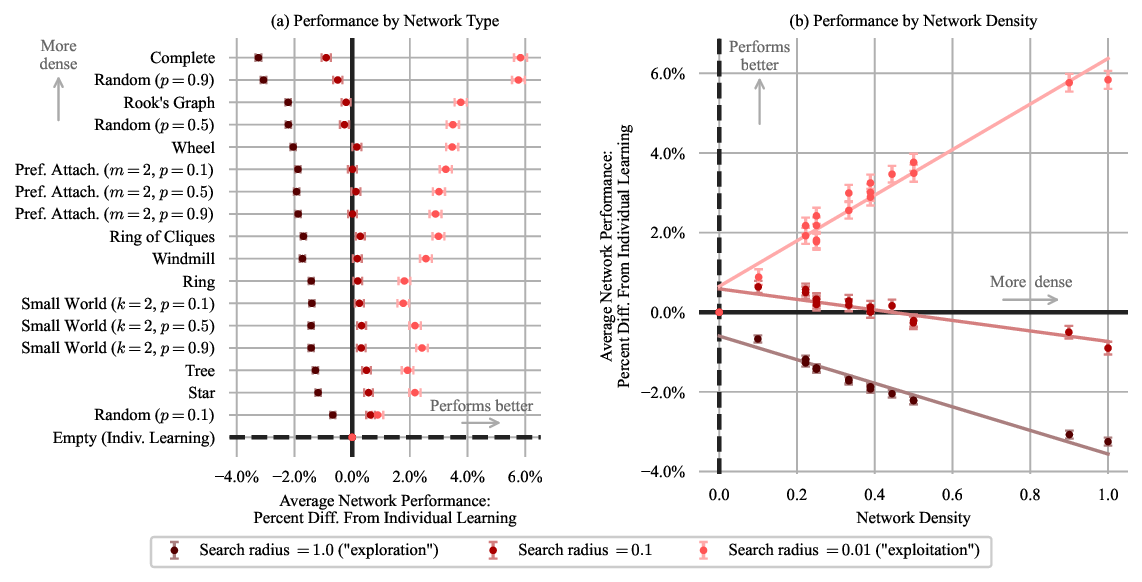}
    \caption{\textbf{Average performances for each network and search radius.} Values shown are percent differences from the averages of teams that only learn individually (empty graphs) for teams of $n=9$ agents with the specified search radius. Error bars show 95\% confidence intervals of the differences between means. (a) Average performances are shown by network type, sorted top-to-bottom first by density, then by eigenvector centrality, a measure of how important each individual is to each other individual. (b) The same performance averages plotted by network density. Trend lines represent simple least squares regressions between density and performance, illustrating how the search radius influences the relationship between network density and performance. SI Appendix S6 shows similar outcomes for teams of 4, 16, and 25.}
    \label{fig:relative_performance}
\end{figure*}

The first result of this analysis is that network density can modestly help or hurt a team's performance depending on how much the team explores new solutions and exploits old ones. Sparse teams tend to outperform dense teams when innovating because of reduced uncertainty, while dense teams outperform sparse teams when refining due to faster optimization (see Fig. \ref{fig:relative_performance} for teams of $n=9$ and three search radii; see SI Appendix S6 for all team sizes and radii). The performance benefits are modest---up to 3.4\% for sparse graphs when exploring and 6.1\% for dense graphs when exploiting.

To build intuition for this finding, recall that agents in teams with dense networks (like the complete graph) have many more connections than in teams with sparse networks. 
Each agent controls its one variable in their neighborhood's performance, and their neighbors' variables can change randomly each turn. 
So if an agent has fewer neighbors, it is more likely that any change the agent makes will produce the expected outcome. 
But the more neighbors an agent has, the more likely it becomes that multiple agents make changes at the same time, landing both agents in a different location on the task landscape than they expected. 
While these changes could be compatible (as they are when agents do not change each other's probabilities), their combined outcome is less certain on tasks tied to network properties.
When exploring globally (radius $=1.0$), agents in dense teams have even more uncertain environments and experience more random changes than agents in teams with fewer connections.
As a result, sparse teams outperform dense teams when innovating.

Next, consider when teams exploit locally (radius $=0.01$). This radius makes agents' changes incremental: changes they make either stay on the same peak in the task landscape or move to an adjacent peak.
As agents adjust toward the summit of their local peak, the probability of jumping to adjacent peaks (if they exist in the task) goes down because fewer points within their search radius perform better than their current solution. 
This is not explicit coordination between agents; much like the optimization method of gradient descent, agents passively ``teach'' each other where good solutions are through small shifts that guide neighbors toward better outcomes for the team.
At the same time, this guidance also performs a passive form of error-checking by preventing other agents from finding solutions with a lower fitness than solutions that are within the current search radius.
So like before, agents in dense teams are exposed to more simultaneous changes than those in sparse teams.
But this time, those changes guide agents with many connections toward local peaks faster than those with few.
Sparsely-connected teams, on the other hand, have to wait for information about the task landscape to pass from agent to agent through their random searches over many turns, leaving them to navigate more slowly toward optimal solutions.
Sparse teams are more likely to refine one dimension at a time, like coordinate descent optimization, while dense teams often refine multi-dimensionally.
Consequently, dense teams outperform sparse teams when exploiting.
Between these extremes, intermediate searches (radius $=0.1$) exhibit mixed results as the shape of each task determines whether density or sparsity yields better outcomes (see SI Appendix S6).

Thus, having fewer connections, sparser layouts, or both makes it easier for multidisciplinary teams to explore novel solutions. Large changes are more likely to improve individuals' contributions in such teams while small changes slow the process of reaching optimal team outcomes. On the other hand, teams with more connections, denser layouts, or both exploit existing solutions more effectively. In this case, small changes help the team optimize performance quickly while large changes create uncertainty throughout the team. This finding is comparable to studies of other types of learning that find the same effect \cite{Lazer2007, Shore2015, Derex2016}. In those contexts, dense networks are ``efficient'' because they spread information and converge to local solutions quickly. The ``inefficiency'' of sparse networks gives individuals greater opportunity to find diverse solutions making them more likely to find better ones \cite{Lazer2007}.

However, density does not account for all---or even most---of the performance variation with mediated learning. In Fig. \ref{fig:relative_performance}a, we sorted the networks from top to bottom first by density. Then, we broke any ties by sorting according to the greatest mean eigenvector centrality, a measure of how central the neighbors of each node are on average \cite{Newman2018}. While these sorts mostly placed the networks in performance order for the global search results, they are not perfectly ordered, let alone the orders of the other search radii. This raises the question of what network qualities best correspond to greater team performance across tasks and search conditions. The next section addresses this question by systematically examining how different task and network qualities relate to performance.

\subsection*{Decentralization improves performance across tasks}
\label{subsec:results-dentralization}

\begin{figure}
    \centering
    \includegraphics[width=0.5\textwidth]{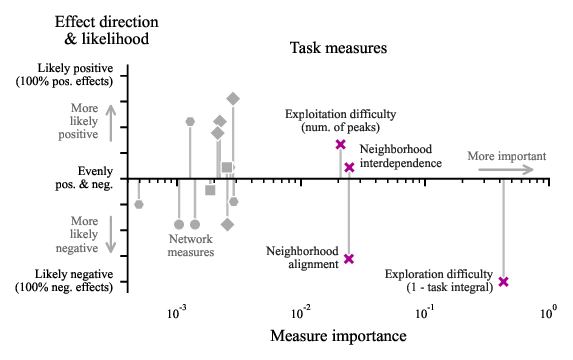}
    \caption{\textbf{Task measure importances, likelihoods, and effect directions.} Each point shows how important the indicated measure is (x-axis), and plots those importances against how likely that measure is to either have a positive effect (positive y-axis) or negative effect (negative y-axis) on team performance. Each point summarizes the fraction of valid regression cumulative effects (combining main and search radii interactions) with a positive effect. For example, exploration difficulty was negative in all 18 cumulative effects, while neighborhood interdependence was positive in 10 of 18 cumulative effects, and negative in 8 of 18.}
    \label{fig:importances_tasks}
\end{figure}

\begin{figure}
    \centering
    \includegraphics[width=0.5\textwidth]{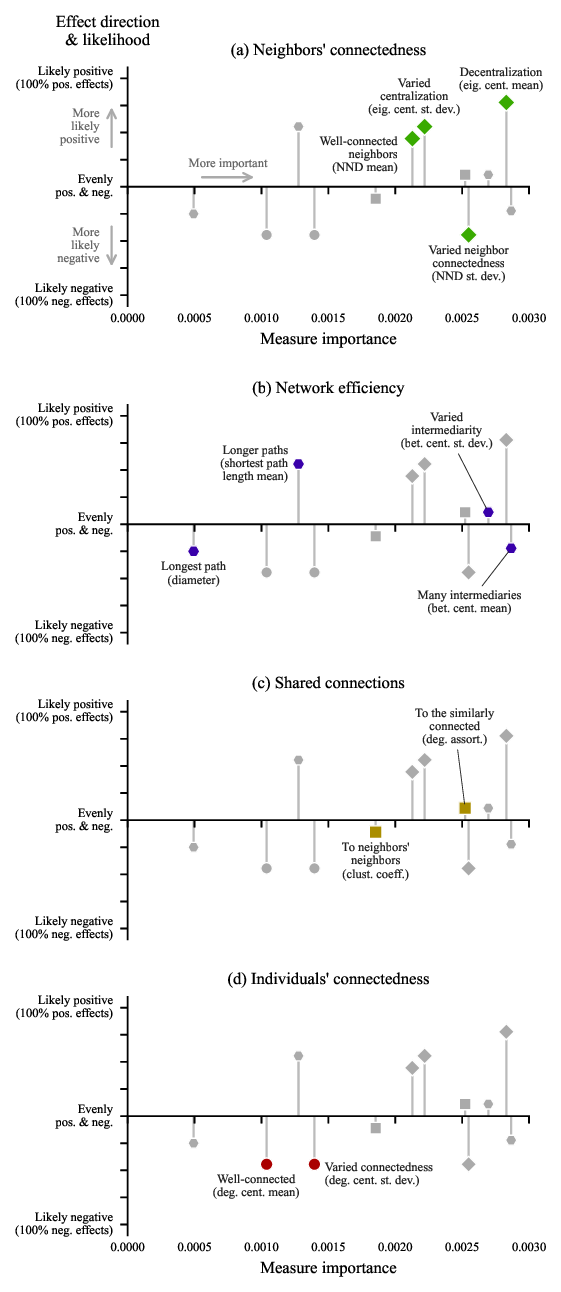}
    \caption{\textbf{Network measure importances, likelihoods, and effect directions.} Subfigures show measures of (a) neighbors' connectedness, (b) network efficiency, (c) shared connections, and (d) individual connectedness. Again, each point shows how important the indicated measure is (x-axis), and plots those importances against how likely that measure is to either have a positive effect (positive y-axis) or negative effect (negative y-axis) on team performance. Decentralization and having many intermediaries are tied for most important (within one standard deviation of each other's means). NND stands for nearest neighbor degree.}
    \label{fig:importances_networks}
\end{figure}

While our analyses have focused on network density so far, we saw that density does not account for most of the performance variation with mediated learning. Therefore, our next two analyses utilized our full dataset (all 18 network types, 34 tasks, four search radii, four team sizes, and 26 time steps for each) to identify the relative contributions of a variety of network qualities to team performance. The first analysis found the relative importances of 12 network measures through random forest regressions  \cite{Breiman2001}. However, random forests cannot measure whether effects are positive or negative. So, our second analysis predicted how likely each measure is to have a positive or negative performance effect by aggregating the results of six regressions with different controls. Figs. \ref{fig:importances_tasks} \& \ref{fig:importances_networks} show our results.

We identified the most important network measures through two random forest regressions (see SI Appendix S7). Each forest included task fixed effects and controls for the four task measures, time step, team size, and search radius. The first forest included the 12 network measures that apply to fully-connected networks. This comprehensive list of network measures came at the cost of excluding data from teams with networks that are not fully connected (e.g. the empty graph, some random graphs) and cannot be evaluated by three of the measures (shortest path length, diameter, and degree assortativity). The second forest excluded network measures that cannot evaluate disconnected graphs, allowing us to include the comprehensive set of all networks regardless of their connectedness. We calculated the average and standard deviation of the importances for each measure across the two random forests to find their impacts on team performance.

Our second analysis extends Mason \& Watts' \cite{Mason2012} application of regressions to assess how well network measures predict performance. However, network measures are often highly correlated and contain similar information to one another \cite{Oldham2019}. To overcome this, we performed six multivariate ordinary least squares regressions, each containing one of six combinations of measures (see SI Appendix S8 for details). We then combined the results of these regressions by calculating how often each measure produced either a positive or negative effect across the regressions. The six combinations consisted of (a) one of three groups of task measures and (b) one of two groups of network measures. For tasks, each regression could include the task measures, task fixed effects, or both task measures and fixed effects. For networks, half of the regressions used completely-connected networks with all network measures, while the other half included all networks but only network measures that are valid on disconnected graphs. All six regressions included controls for time step, team size, the square of team size, and search radius. Each regression also included search radius interaction effects for task and network measures as other studies have found that task and search objectives can moderate desirable qualities \cite{Shore2015, Amelkin2018}. The regressions gave us main effects and search radius interaction effects for task and network measures. Finally, we counted how often each measure yielded a positive cumulative effect. Each cumulative effect was built by adding the main and interaction effects for the exploration, exploitation, and mixed search radii (radius$=$0.01, 0.1, and 1.0) whenever they were statistically significant. This gave estimates of how likely each measure's effect is to be positive or negative.

The combined results of these analyses show several items of note. First, task qualities likely affect performance more than network measures, by one to two orders of magnitude (Fig. \ref{fig:importances_tasks}). Exploration difficulty was both the most likely measure to affect performance and the most important, followed by alignment between individuals' task neighborhoods (their objectives) which is tied as the second-most likely effect. Together, task qualities accounted for about 50.0\% of team performance effects. Even if saying ``hard tasks are hard'' is not surprising, it reaffirms prior findings on task complexity's moderating effect \cite{Barkoczi2016, Almaatouq2021}.

Network measures accounted for about 2.4\% of team performance effects, which is 4.8\% the size of the task effects. The most important group of network measures was neighbors' connectedness (Fig. \ref{fig:importances_networks}a). Decentralization is measured through the mean of eigenvector centrality which has a normalized euclidean norm, meaning its average is maximal in all cases where nodes are equally central (true from empty to complete graphs, see SI Appendix S9). That measure is tied for the most important network measure (within one standard deviation of mean betweenness centrality, a measure of intermediarity). It also is the most likely network effect, the most likely positive effect, and is tied with alignment as the second most likely measure overall. The other measures of neighbors' connectedness (including varied centralization) fill four of the top seven network measures revealing a consistent importance and likelihood not seen in the other groups. While it may seem contradictory for variation measures to score highly alongside decentralization, it need not be. Varied centralization is measured through the standard deviation of eigenvector centrality. Networks like the ring of cliques, the wheel, and the windmill simultaneously exhibit moderate decentralization and high variance in centralization (see SI Appendix S9). These networks performed modestly across most tasks demonstrating how teams can possess both qualities simultaneously.

Results for the other measures proved mixed. Network efficiency---having short paths between individuals---often proves important for other types of learning \cite{Lazer2007, Mason2008, Mason2012, Barkoczi2016}. For mediated learning, we find that efficient networks have mixed effects (Fig. \ref{fig:importances_networks}b). Having many intermediaries was tied for the most important network measure, and varied intermediarity throughout a team came in third. However, these measures did not consistently produce positive or negative effects. Given their importances, this suggests that efficiency still affects performance, but likely through higher-order interactions. Longer paths resulted in more consistent positive gains but with less importance, while measures of shared connections (Fig. \ref{fig:importances_networks}c) were somewhat important but with inconclusive directionality. Last, high levels of individual connectedness and variation therein produced moderately consistent negative effects but with less import (Fig. \ref{fig:importances_networks}d).

In sum, with mediated learning, decentralization is the most important and most likely network quality to improve multidisciplinary team performance across the data from all 34 tasks, followed by other qualities of network efficiency and neighbors' connectedness in line with previous studies.

\section*{Discussion}
\label{sec:discussion}


Many teams divide tasks into separate, interrelated subtasks by discipline, much like how cooks collaborate in a high-volume restaurant kitchen \cite{Rambourg2013, Gregoire2016}. In this study, we explored how members of such teams learn from one another mediated by their neighborhood performances even when they cannot see each other's actions or outcomes. We found that a team's interaction network can significantly affect team performance, but only on tasks that explicitly rely on network properties. The effects of network qualities are relatively small at about 5\% the effect size of task qualities. Still, this proved sufficient to give sparsely-connected teams an advantage (up to 3.4\%) when innovating and densely-connected teams an advantage (up to 6.1\%) when refining their solutions across our set of 34 tasks. Such advantages could prove consequential in contexts where modest performance gains provide an edge over competitors, for example, but may be too costly to implement in contexts that necessitate sizeable improvements.

Furthermore, we found that decentralizing a team's network by evenly distributing coordination responsibilities among a team's members tends to improve performance across our battery of 34 tasks. As with other types of learning \cite{Lazer2007, Fang2010, Mason2012, Barkoczi2016, Brackbill2020}, the efficiency with which a network distributes information appears likely to affect performance, but our analyses yielded mixed results on the importance of network efficiency and other qualities of networks.


\begin{figure}
    \centering
    \includegraphics[width=0.48\textwidth]{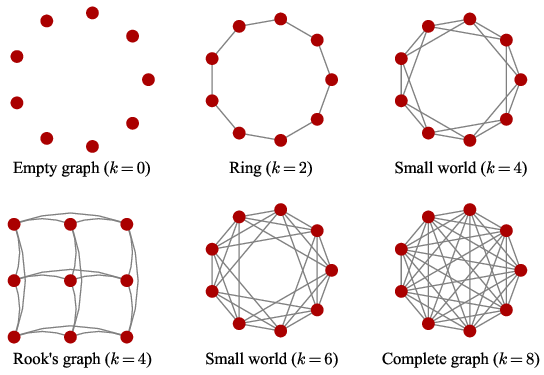}
    \caption{\textbf{Examples of decentralized networks.} The most-decentralized networks tend to evenly assign the same number of connections $k$ to each node, as one might evenly distribute coordination responsibilities among members of a team.}
    \label{fig:decentralized_networks}
\end{figure}

Our results provide insights into better ways for multidisciplinary teams to coordinate. Given mediated learning, teams would benefit from carefully shaping their search objectives, interaction networks, and tasks---depending on what they can control at the time---as each holds repercussions for their performance. For example, when teams can control their search conditions, teams may be more effective at refining already-complex products through incremental changes than by letting individuals run wild (Fig. \ref{fig:relative_performance}). Likewise, it may be easier for one or two individuals to explore new solutions on their own than for the whole team to equivocate together.

When teams can control their network, moving away from centralized formal structures and toward decentralized ones (Fig. \ref{fig:decentralized_networks}) may improve team outcomes across a variety of tasks, similar to other findings on social networks \cite{Rulke2000, Becker2017, Xu2022}. This may be especially important early in team formation because interaction networks are often rigid, tending toward fixed formal structures due to lower coordination costs along those pathways \cite{Clement2018}. But this may not mean that teams should self-organize their structures, as even weakly-enforced formal structures tend to out-compete such teams \cite{Clement2018}. While much remains unknown, our findings indicate that teams are likely to benefit from formally coordinating decentralized patterns of interaction to improve their performance.

When networks are difficult to control or already in place, restructuring tasks could improve performance. Task qualities, like difficulty and interdependence, may affect team performance by one to two orders of magnitude more than network qualities (Fig. \ref{fig:importances_tasks}). This presents opportunities to explore methods of improving performance and reducing uncertainty through task redesign---by making task landscapes less difficult for search conditions (e.g. innovation, refinement), reshaping subtasks at strategic locations throughout the network, at particular times in an iterative search process, etc. Altogether, our findings support the growing consensus around the necessity of adaptive team learning for optimal performance \cite{Levinthal1997, Burke2006, Baard2014, Christian2017, Yang2021, Almaatouq2021}.

Our findings also suggest ways that organizations may improve their performance. Scholars have long applied the methods we use here to study questions of organization design \cite{March1991, Lazer2007, Rivkin2007, Fang2010, Clement2018} as well as teams. Limiting a department's connections to outside departments may aid innovation, supported by famous design shop successes like the Lockheed ``Skunk Works'' \cite{Johnson1985}. Likewise, gradual changes may help organizations improve complex products more readily than making large changes \cite{Helfat2011}. And importantly, decentralizing coordination activities may improve performance more than centralizing coordination \cite{Matusik2021, Venkataramani2023}.

Our evidence also supports that tasks can limit a team's collective intelligence \cite{Graf-Drasch2021}. Mediated learning could underlie parts of team learning wherever knowledge boundaries exist within groups. For example, some teams in studies of collective intelligence divide up tasks and specialize \cite[e.g.][]{Amelkin2018} making our findings relevant to those tasks at least. We do not mean to suggest that testing teams with tasks that limit collective intelligence makes those tasks invalid; to the contrary, those tasks provide teams with opportunities to demonstrate their collective intelligence by adapting to the tasks at hand \cite{Almaatouq2020}. Communication patterns that arise within teams---whether as the product of social perceptiveness \cite{Woolley2010, Riedl2021}, virtuality \cite{Engel2014, Engel2015}, or homophily \cite{McPherson2001}---may influence a team's collective intelligence through current task and network constructions. Thus, our findings support Graf-Drasch et al.'s \cite{Graf-Drasch2021} recommendation that we re-conceive of collective intelligence as a multi-dimensional concept like we do with individual intelligence, including a team's ability to adjust their interaction pattern to the task.

Finally, our results may extend to performance-related networks with non-additive effects in adjacent fields such as population genetics. For example, in some contexts, our finding that dense networks exploit better may offer an alternative search dynamic to shifting balance theory, which highlights the way in which adaptive search can be more efficient on sparser epistatic networks \cite{Wright1982}.


Just as this work conceptually benefits from simplifying qualities like individual skill and social learning, its implications are also limited by these assumptions. Our findings do not suggest how introducing varied skills or individual intelligence may moderate these results, nor how any copying or coordination that takes place in multidisciplinary teams may influence them. For example, we performed an experiment with slightly increased coordination between agents, an alternative approach to minimizing skills by granting a small amount of coordination (a form of team learning) while removing the small individual experimentation skill. In this case, we do not see strong positive effects of mediated learning, though we do see significant negative effects (see SI Appendix S10, Fig. S17). The relative importances of network properties also shifted in this case. While decentralization was still an important positive performer, intermediaries and clustering proved more important, albeit with less confidence in their positive effects (Fig. S20). Hence, more research on team process variables of all kinds---including different search strategies---remains crucial for understanding team performance.

Our work is also limited by our selection of tasks, networks, and network measures. We sought to select tasks, networks, and measures that covered qualities of broad interest and rapport. Still, more nuanced structures and measures would aid identification of network qualities that perform well, both on specific tasks and across many tasks. Our regression analyses raised as many questions as they answered about the qualities that shape performance in the context of mediated learning. Network features interact and cannot be fully independent from one another. The result is a complex system whose output is never fully governed by one feature or mechanism alone, leaving the subject ripe for future examination.

\matmethods{
\subsection*{Model pseudocode}
\begin{algorithmic}[1]
\Eeee{Graph with vertices $i \in I$, neighbors $j \in J_i$ $\forall$ $i \in I$, neighborhood objective functions $g_i(x_i,\Vec{x}_j)$ $\forall$ $i \in I$ where $\Vec{x}_j = (x_{j=1},\ldots,x_{j=k_i})$, graph objective function $h(\Vec{x})$ where $\Vec{x} = (x_{i=1},\ldots,x_{i=n})$, and initial positions $x_i$ $\forall$ $i \in I$.}
\Oooo{Performance values $g_i$ $\forall$ $i \in I$ and $h$.}
\FORALL{timestep}
    \FORALL{$i \in I$}
        \STATE{Draw $\Delta x$ from $\mathcal{U}(-\epsilon,\epsilon)$}
        \STATE{$x'_i$ := $x_i + \Delta x$}
        \STATE{$g_i$ := $g_i(x_i,\Vec{x}_j)$}
        \STATE{$g'_i$ := $g_i(x'_i,\Vec{x}_j)$}
    \ENDFOR
    \FORALL{$i \in I$}
        \IF{$g'_i>g_i$}
            \STATE{$x_i$ := $x'_i$}
        \ENDIF
    \ENDFOR
    \STATE{$h=h(\Vec{x})$}
\ENDFOR
\end{algorithmic}

\subsection*{Code \& data availability}

The model and analysis code for this work were developed with Python 3.9.5, Numpy 1.21.6, Scipy 1.9.0, Pandas 1.4.2, NetworkX 2.8.3, Dask 2022.4.1, Statsmodels 0.13.2, Scikit-learn 1.0.2, Matplotlib 3.5.2, and Seaborn 0.11.2. Complete code is available at  \href{https://github.com/meluso/multidisciplinary-learning}{https://github.com/meluso/multidisciplinary-learning}. Data for this work is available at \href{https://osf.io/kyvtd/?view_only=5cb48a75acbd445caac377fce17f6e1c}{https://osf.io/kyvtd}.
}

\showmatmethods{} 

\acknow{Our sincere thanks to James Bagrow for his time and thoughts, particularly his suggestions to include the rook's graph and the empty graph as a baseline. Special thanks to Mirta Galesic and the reviewers for manuscript feedback. Thanks also to Jean-Gabriel Young, Melania Lavric, Kendall Fortney, Babak Heydari, and the many others who supported us on this work.}

\showacknow{} 

\bibsplit[11]

\bibliography{library.bib}

\end{document}